\documentclass[prb,aps,amssymb,shownopacs,twocolumn,10pt]{revtex4-1}

\usepackage{amsmath}
\usepackage{amssymb}
\usepackage{amsthm}
\usepackage{amsfonts}
\usepackage{listings}
\usepackage{easybmat}
\usepackage{enumerate}
\usepackage{latexsym}
\usepackage{epstopdf}
\usepackage{psfrag}
\usepackage{color}
\usepackage{bm}
\usepackage[pdftex]{graphicx}
\usepackage{hyperref}
\usepackage{ulem} % to strike things out 
\newcommand{\moire} {moir{\' e} }

\newcommand{\beq}{\begin{equation}}
\newcommand{\eneq}{\end{equation}}
\def\be{\begin{equation}}
\def\ee{\end{equation}}
\def\ba{\begin{eqnarray}}
\def\ea{\end{eqnarray}}

\def\Tr{{\rm Tr}}

\input{epsf}

\begin{document}

\title{
Chern bands of twisted bilayer graphene:\\ fractional Chern insulators and spin phase transition
}

\author{C\'ecile Repellin$^{1, 2}$}
\author{T. Senthil$^2$}
\affiliation{$^1$ Univ. Grenoble-Alpes, CNRS, LPMMC, 38000 Grenoble, France\\
$^2$ Department of Physics, Massachusetts Institute of Technology, Cambridge, MA, 02139}

\date{\today}
\begin{abstract}
When one of the graphene layers of Magic Angle Twisted Bilayer Graphene is nearly  aligned with its hexagonal boron nitride substrate (a configuration dubbed TBG/hBN), the active electronic bands are nearly flat, and have a Chern number $C=\pm1$. Recent experiments demonstrated a quantum anomalous Hall effect and spontaneous valley polarization at integer filling $\nu_T=3$ of the conduction band in this system. Motivated by this discovery, we ask whether fractional quantum anomalous Hall states (FQAH) could also emerge in TBG/hBN. We focus on the range of filling fractions where valley ferromagnetism was observed experimentally. Using exact diagonalization, we find that the ground states at $\nu_T = \frac{10}{3}$ and $\nu_T=\frac{17}{5}$ are fractional Chern insulator states in the flat band limit (in the hole picture, these are the fractional quantum Hall fractions $\frac{2}{3}$ and $\frac{3}{5}$).  The ground state is either spin polarized or a spin singlet depending sensitively on band parameters. For nominally realistic band parameters, spin polarization is favored.  Flattening the Berry curvature by changing a band parameter gives way to the spin singlet phase.  Our estimation of the charge gap in the flat band limit shows that the FQAH state may be seen at accessible temperatures in experiments. We also study the effect of a non-zero bandwidth and show that there is a reasonable range of parameters in which the FQAH state is the ground state.
\end{abstract}

\maketitle

The  nearly flat bands of many moir\'e graphene materials are a platform for several fascinating many-body phenomena, including correlated insulators~\cite{cao2018correlated,cao2018unconventional,Wang2019Evidence,yankowitz2019tuning,lu2019superconductors, Shen2019Observation,Liu2019Spin, Cao2019Electric}, superconductivity~\cite{cao2018unconventional,yankowitz2019tuning,lu2019superconductors,Wang2019Signatures,Liu2019Spin,Shen2019Observation,stepanov2019interplay,saito2019decoupling}, ferromagnetism~\cite{Sharpe_2019},\cite{Liu2019Spin, Shen2019Observation,Cao2019Electric, chen2019tunable,serlin2019intrinsic}, and a (quantized) anomalous Hall effect~\cite{Sharpe_2019, serlin2019intrinsic, chen2019tunable}. 
Theoretically it has also become clear that in many moir\'e graphene materials, these nearly flat bands are also topologically non-trivial~\cite{po-PhysRevX.8.031089, zhang2019nearly, song-PhysRevLett.123.036401, po-PhysRevB.99.195455, ahn-PhysRevX.9.021013, bultinck2019anomalous, zhang2019twisted, liu-PhysRevX.9.031021}. 
Thus these systems provide an experimental context where the interplay between band topology and electron correlations must be confronted.  

Of particular interest to us is magic angle twisted bilayer graphene where one of the graphene layers is nearly aligned with a hexagonal boron nitride (h-BN) substrate. In this system (denoted TBG/hBN), a single-particle gap separates the valence and conduction bands at charge neutrality~\cite{zhang2019twisted, bultinck2019anomalous}. In each of these bands, time reversal relates one valley to the other so that the Chern number in either valley is $\pm 1$\footnote{The moir\'e potential decouples the two valleys from each other in the band structure so that the Chern number of each valley-filtered band is well defined.}. Note that these properties are absent from {\it unaligned} magic angle TBG (a case beyond the scope of this paper), which is characterized by 'fragile' topology~\cite{po-PhysRevX.8.031089, po-PhysRevB.99.195455, song-PhysRevLett.123.036401, ahn-PhysRevX.9.021013}, not by a Chern number. Experimentally, emergent ferromagnetism was observed in TBG/hBN~\cite{Sharpe_2019, serlin2019intrinsic} at filling $\nu_T=3$ of the conduction band (including spin and valley degrees of freedom). It is accompanied by a large~\cite{Sharpe_2019} or even quantized~\cite{serlin2019intrinsic} anomalous Hall effect. 
This result is most simply understood in the hole picture, where the hole filling is $\nu_{h,T} = 4-\nu_T$. The band insulator at $\nu_{h,T} = 0$ has no net Chern number; at $\nu_{h,T} = 1$, in a spin and valley polarized ground state, the holes occupy a single Chern band leading to a quantized anomalous Hall (QAH) state.
Ferromagnetism and QAH effect were already predicted~\cite{zhang2019nearly} in the closely related context of multilayer moir\'e graphene materials in the limit where the interaction strength far exceeds the bandwidth. The theoretical understanding of ferromagnetism in narrow bands of \moire systems has since expanded to include evidence from Hartree-Fock calculations~\cite{lee2019theory,bultinck2019anomalous, zhang2019bridging, zhang2019twisted, xie2018nature, alavirad-2019arXiv190713633A}, as well as exact diagonalization and density matrix renormalization group (DMRG) calculations~\cite{repellin-2019arXiv190711723R}.
% Ferromagnetism and quantized anomalous Hall effect have also been reported in experiments on other \moire graphene systems~\cite{}.

In this context, it is of tremendous interest to ask if TBG/hBN also shows a Fractional Quantum Anomalous Hall (FQAH) effect when the conduction band is doped away from integer filing. Fractional quantum Hall (FQH) states in lattice systems are known as fractional Chern insulators (FCI). 
There is an extensive prior literature on FCI states in toy models of interacting electrons in narrow Chern bands. For a single component system, the existence of FCIs in theoretical models has been demonstrated from analytical and numerical calculations~\cite{neupert-PhysRevLett.106.236804, sheng-natcommun.2.389, regnault-PhysRevX.1.021014} (see also Refs.~\onlinecite{Parameswaran_2013, BERGHOLTZ_2013} and references within). An understanding  of the conditions facilitating their emergence appears from this literature; in analogy to Landau Level (LL) physics, a nearly energetically flat Chern band with a nearly flat Berry curvature and quantum metric provides a favorable platform for FCIs~\cite{wu-PhysRevB.85.075116, parameswaran-PhysRevB.85.241308, Parameswaran_2013, grushin-PhysRevB.86.205125, BERGHOLTZ_2013, bauer-PhysRevB.93.235133, jackson-natcomm2015}.
As a nearly flat Chern band is already realized in TBG/hBN, it may be a good platform for the FQAH effect.
Though FCIs are already realized in (AB stacked, not twisted) bilayer graphene aligned with hBN in the presence of a magnetic field~\cite{Spanton_2018}, here we are concerned with states that occur in zero external magnetic field. 
The band of TBG/hBN has a non-trivial Berry curvature distribution and is not ``Berry-flat"; a detailed calculation is thus needed to address the possibility of an FCI .

Multicomponent FCIs have not been studied much, unlike their single-component couterparts. The literature in this case is mostly limited to two-component models preserving time-reversal symmetry~\cite{Neupert_2011, ghaemi-PhysRevLett.108.266801, repellin-PhysRevB.90.245401}.
Generically for multicomponent Chern bands, in the limit of a flat band with uniform quantum geometry, we may expect states connected to multicomponent FQH physics in a LL. Examples of such physics include multicomponent (Halperin) model wavefunctions~\cite{Halperin83}, and fractionally charged excitations with an extended spin texture (skyrmions)~\cite{sondhi-PhysRevB.47.16419}. However, it is unknown how these multicomponent phenomena are influenced by the band structure properties relevant in FCIs, such as non-zero bandwidth and non-trivial Berry curvature distribution.

With these motivations we study a model appropriate for TBG/h-BN and present evidence for a FQAH state at total band fillings $\nu_T = \frac{10}{3}$ and $\nu_T=\frac{17}{5}$, {\em i.e} a filling per spin per valley of $\frac{5}{6}$ or $\frac{17}{20}$ (equivalent to a hole filling $\nu_{h,T} = \frac{2}{3}$ and $\nu_{h,T} = \frac{3}{5}$). Ref.~\onlinecite{serlin2019intrinsic} reports hysteresis in a range of fillings that includes these values of $\nu_T$. We interpret this as evidence for valley polarization, and use a valley-polarized model throughout the paper. We thus have holes in a band with Chern number $C = 1$ at $\nu_{h,T} = \frac{2}{3}, \frac{3}{5}$. Existing experiments do not directly give evidence for spin polarization; thus we do not assume any spin polarization.
Nevertheless, and in contrast to the fate of electrons in a Landau level at the same filling~\cite{eisenstein-PhysRevB.41.7910, shayegan-PhysRevB.45.3418, kukushkin-PhysRevLett.82.3665, Yacoby-natPhys2007, wu-PhysRevLett.71.153, wu-PhysRevB.49.7515}, we find that, for nominally realistic band parameters, the ground state at $\nu_{h,T} = \frac{2}{3}$ is spontaneously spin polarized in the strong interaction limit, and is a FCI state. The topological properties of this state are the same as the particle-hole conjugate~\footnote{Here, we refer to the particle-hole conjugation which acts in the spin-polarized Hilbert space, transforming a state at $\nu$ to one at $1-\nu$} of the $1/3$ Laughlin state. As the parameters of the underlying band structure are changed we demonstrate a phase transition to the spin unpolarized Halperin (112) state~\cite{Halperin83}. 
Similarly at $\nu_{h,T} = \frac{3}{5}$ we find a spin-polarized FCI (akin to the particle-hole conjugate of the $2/5$ Jain state), and a transition to a spin-unpolarized state.
Generically, we find that larger fluctuations of the Berry curvature stabilize the ferromagnet compared to the spin singlet. They may also destroy the fractional quantum Hall phase altogether and lead to a spin-polarized metal in the flat-band limit. 
Based on calculations of the charge gap in this limit we show that the FQAH state may be seen at accessible temperatures in experiments.  We also study the effect of a non-zero bandwidth and show that there is a reasonable range of parameters in which the FQAH state is the ground state.

\begin{figure}
\includegraphics[width=0.55\linewidth]{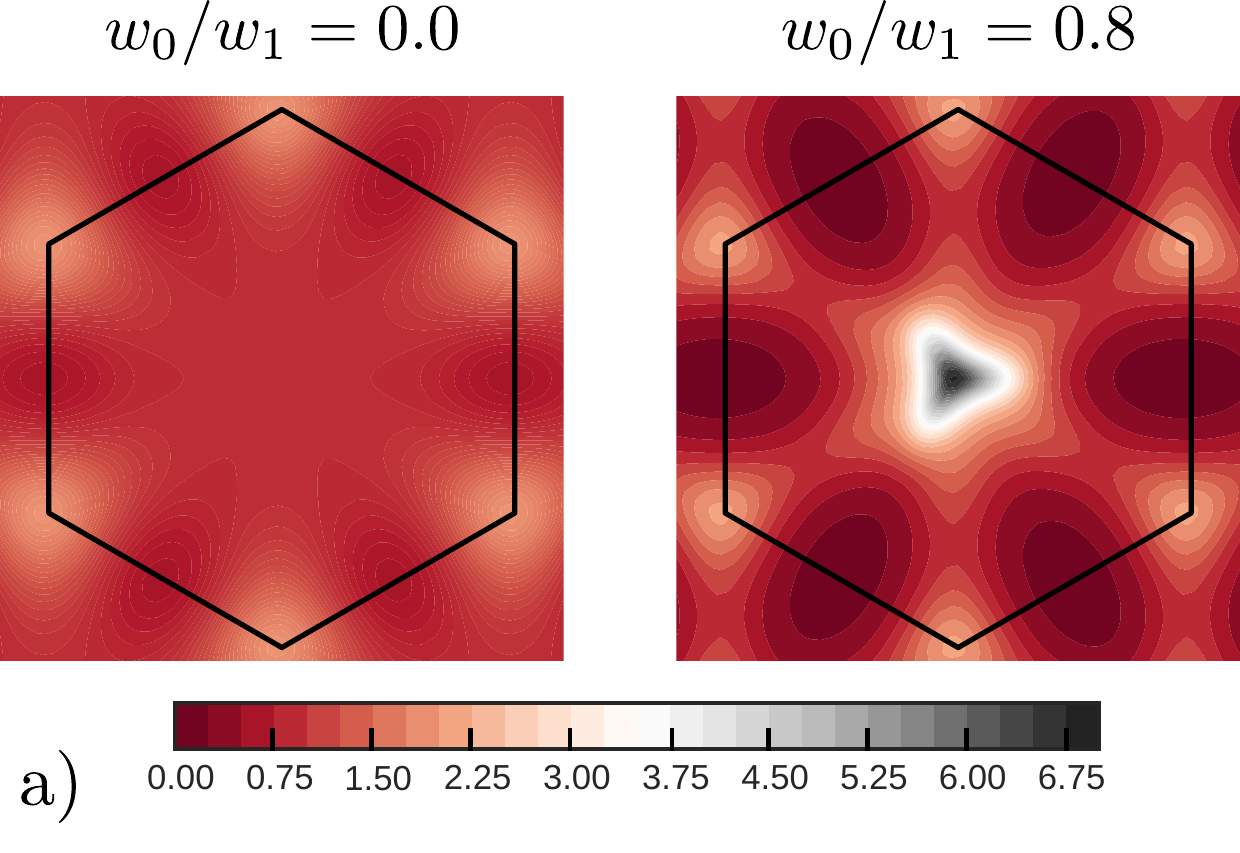}
\includegraphics[width=0.38\linewidth]{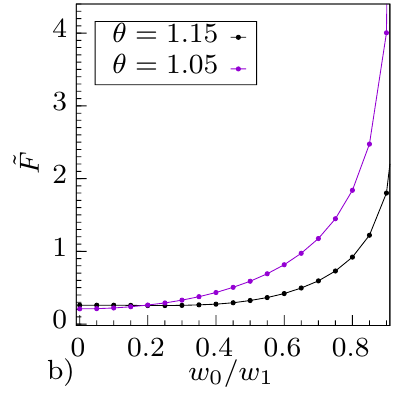}
\caption{a) Distribution of the Berry curvature density of the conduction band of TBG/hBN across the Brillouin zone for $w_0/w_1 = 0.0$ and $w_0/w_1 = 0.8$. b) Evolution of the root mean square of the Berry curvature distribution with the relaxation parameter $w_0/w_1$, for two values of the twist angle $\theta = 1.05$ and $\theta = 1.15$. In the rest of the calculations, we fix $\theta = 1.15$.}
\label{fig: single particle}
\end{figure}

\textit{Model}
We consider the continuum momentum-space model~\cite{bistritzer2011moire} of TBG/hBN~\cite{zhang2019twisted, bultinck2019anomalous} (for concreteness we choose the twist angle $\theta = 1.15^{\circ}$) and assume valley polarization. Due to the breaking of sublattice symmetry in this model, valence and conduction band are separated by an energy gap, and the conduction band has a Chern number $C=1$.
We take the limit where it is separated from other bands by a gap much larger than its bandwidth.
The many-body Hamiltonian is obtained by projecting the screened Coulomb interaction $V(q)$ to the conduction band:
\begin{eqnarray}
H_V  & = &\sum_{\mathbf q} : \tilde{\rho}(\mathbf q) V(\mathbf q)\tilde{\rho}(-\mathbf{q}): \\ 
V(q) & = & U\frac{1}{q}\left(1 - e^{-qr_0} \right)
\label{eq: screened coulomb}
\end{eqnarray}
$\tilde{\rho}(\mathbf q) = \tilde{\rho}_{\uparrow}(\mathbf q) + \tilde{\rho}_{\downarrow}(\mathbf q)$ is the spinful density operator projected onto the conduction band, and $r_0$ is the screening length (we choose $r_0= 5.0$ in units of the \moire lattice constant). $U$ is roughly estimated\cite{cao2018correlated} to be $\approx 20 meV$ but may possibly be larger\cite{Xie:2019aa}.

To obtain a realistic model of TBG/hBN, we have considered the value $0.8$ for the relaxation parameter $w_0/w_1$, the ratio of interlayer tunneling amplitudes between AA ($w_0$) and AB ($w_1$) sites (we take $w_1 = 110meV$). For these parameters, the conduction band is characterized by a Berry curvature with significant fluctuations (see Fig.~\ref{fig: single particle}a). Such fluctuations may in principle hinder the emergence of a FCI phase even in the flat band limit~\cite{wu-PhysRevB.85.075116, parameswaran-PhysRevB.85.241308, Parameswaran_2013, grushin-PhysRevB.86.205125, bauer-PhysRevB.93.235133, jackson-natcomm2015}.
For $w_0/w_1 = 0$, Ref~\onlinecite{tarnopolsky-PhysRevLett.122.106405} showed the existence of a chiral symmetry which guarantees the perfect flatness of the band at magic angle, as well as exact analytical properties for the single-particle wave functions. An important feature of the limit $w_0/w_1 = 0$ is that the Berry curvature fluctuations are very small close to magic angle (see Fig.~\ref{fig: single particle}a). To study the role of the Berry curvature distribution on the many-body properties, we use $w_0/w_1$ as a control parameter (the conduction band remains gapped for all the considered values of $w_0/w_1$).

We study the many-body phase diagram of TBG/hBN for $0\leq w_0/w_1 < 0.95$ using exact diagonalization. 
We hole-dope the conduction band with a fraction $\nu_{h,T} = N/N_s$ (where $N$ is the number of holes and $N_s$ is the number of $\mathbf{k}$ points in the Brillouin zone); the total filling fraction including spin and valley degrees of freedom is $\nu_T= 4 - \nu_{h,T}$
We focus on $\nu_{h,T}=\frac{2}{3}$ and $\nu_{h,T}=\frac{3}{5}$; because of the Chern number $C=1$ of the band, $\nu_{h,T}$ is the relevant filling for comparisons with LL physics.
% While the lower values of $w_0/w_1$ may not be realized in experiments, we will show that they provide important physical insight. 
We emphasize that the Berry curvature distribution is highly sensitive to microscopic parameters such as twist angle and relaxation parameter (see Fig.~\ref{fig: single particle}c). 
Other factors such as interaction-induced band corrections and relaxation disorder may also renormalize it. Given the important variations of microscopic parameters in TBG samples, it is likely that a rich phenomenology of quantum geometries may be realized now or in the future. It is thus important to gain an understanding of strong interactions in a spinful Chern band in a realistic regime of parameters and beyond.\\

\begin{figure}
\includegraphics[width=0.98\linewidth]{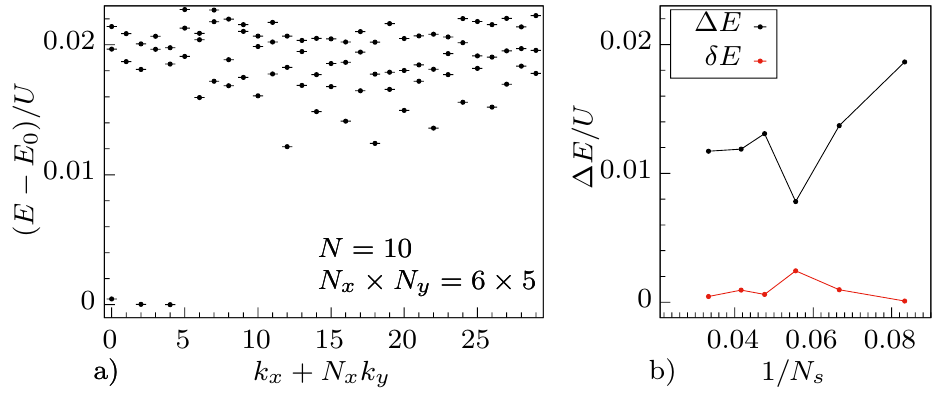}
\caption{Numerical evidence for a FCI ferromagnet at $\nu_{h,T} = 2/3$. a) Low-energy spectrum of the system with $N=10$ fermions and $N_s = 30$ moir\'e sites, for $w_0/w_1 = 0.8$, in the spin-polarized sector. b) Finite-size extrapolation of the many-body gap $\Delta E$ and ground state degeneracy splitting $\delta E$ for $w_0/w_1 = 0.8$.}
\label{fig: FM FCI}
\end{figure}

\textit{Evidence for a ferromagnetic FCI state at $w_0/w_1 = 0.8$}
We start by showing evidence for a FQAH state at $\nu_{h,T}=\frac{2}{3}$ and $w_0/w_1 = 0.8$ in the flat band limit.
For all the system sizes we studied (up to $N=12$ for the spinful system), we observe a full spin polarization.
We identify the ferromagnetic ground state as a FCI akin to the particle-hole conjugate of the $\nu=1/3$ Laughlin state. The low-energy spectrum (see Fig.~\ref{fig: FM FCI}a) indeed consists of three nearly degenerate ground states separated by a gap $\Delta E$ from higher energy excitations. The momentum quantum numbers of these three states are those expected in FQH physics using a mapping of the torus momentum in the continuum to the lattice Brillouin zone~\cite{bernevig-PhysRevB.85.075128}. Gathering the data from several system sizes (up to $N = 20$), $\Delta E$ extrapolates to around $0.01 U$ in the thermodynamic limit. The degeneracy splitting $\delta E$ of the ground state manifold is small ($\delta E < 0.002$), we expect that it will vanish exponentially in the thermodynamic limit (see Fig.~\ref{fig: FM FCI}b). We notice a lower value of $\Delta E$ when the number of k points $N_s$ is a multiple of 9 (for $N_s = 27$, there is even no clear degeneracy at $w_0/w_1=0.8$, although it is present for smaller values of $w_0/w_1$). We interpret it as a possible competition to a charge ordered phase favored by a commensurate geometry; given the stability of $\Delta E$ for all other geometries, we expect the FCI to be the thermodynamic limit ground state.

To confirm the topological nature of the ferromagnetic ground state, we have extracted the nature of the bulk quasihole excitations by calculating a particle entanglement spectrum (PES)~\cite{sterdyniak-PhysRevLett.106.100405, haldane-PhysRevLett.67.937}. The PES is the spectrum of $-\log \rho_A$ where $\rho_A = \Tr_B \rho$ is the reduced density matrix obtained by tracing over $N_B \equiv N - N_A$ particles.
Generically for FQH states, the PES has low levels separated from higher levels by an entanglement gap~\cite{sterdyniak-PhysRevLett.106.100405}. The number of levels below the gap is related to the number of quasihole states in the system with $N_A$ particles and $N_s$ flux quanta; it is a fingerprint of a given topological order.
In particular it can distinguish FCIs from competing phases (such as charge density waves)~\cite{regnault-PhysRevX.1.021014}.
We have calculated the PES, after performing a particle-hole conjugation of the three nearly degenerate ground states obtained from exact diagonalization. The PES is gapped and the number of states below the entanglement gap matches the expectation for a Laughlin $1/3$ state. For example, starting from the ground state of the system with $N=20$ particles on a $6\times 5$ lattice, we find 23256 states below the entanglement gap in the $N_{A} = 5$ partition, in agreement with the expected degeneracy of quasihole states in the FQH system with $5$ fermions and $30$ flux quanta~\cite{bernevig-PhysRevB.85.075128, haldane-PhysRevLett.67.937} (see Appendix~A for further details).

We find similar results at $\nu_{h,T}=\frac{3}{5}$. The ground state is fully polarized for $w_0/w_1 = 0.8$ and it has an approximate fivefold degeneracy with a gap $\Delta E \simeq 0.005U$ to neutral excitations. The PES confirms that it has the same topological order as the particle-hole conjugate of the Jain $2/5$ state~\cite{jain-PhysRevLett.63.199, liu-PhysRevB.87.205136, lauchli-PhysRevLett.111.126802} (see Appendix~B for further details).

\begin{figure}
\includegraphics[width = 0.98\linewidth]{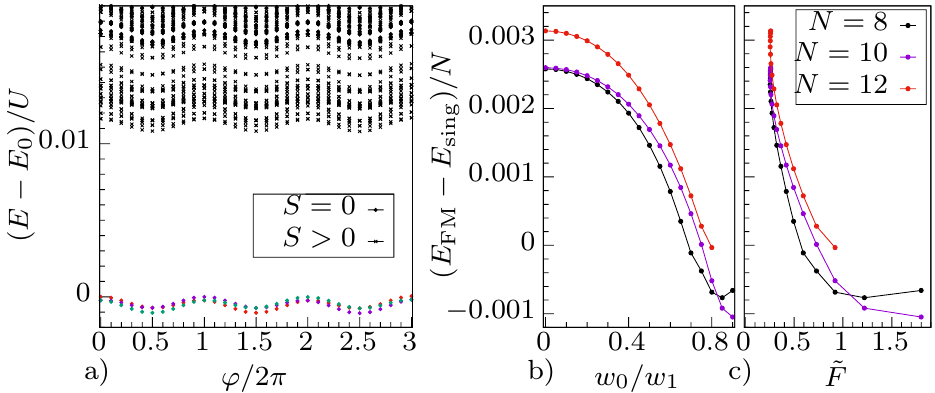}
\caption{a) Low-energy spectrum of the system with $N=10$ spinful fermions and $N_s = 15$ unit cells, for $w_0/w_1 = 0.0$, upon insertion of a magnetic flux $\varphi$. The three topological sectors (red, green and purple dots) have a spectral flow consistent with the topological order of the Halperin (112) state. Both the spin singlet states ($S = 0$) and the other spin sectors are represented. b),c) Energy difference between the singlet ground state and the fully polarized ground state b) as a function of the relaxation parameter $w_0/w_1$ c) as a function of the root mean square $\tilde{F}$ of the Berry curvature distribution.}
\label{fig: singlet}
\end{figure}

\textit{Spin polarization and role of the quantum geometry}
The emergence of a {\it ferromagnetic} FCI state at $\nu_{h,T} = 2/3$ and $3/5$ in the absence of any Zeeman field $H$ is surprising. Indeed, in QH systems involving Landau levels rather than Chern bands, the FQH state observed at these filling fractions has no spin polarization at small $H$, and undergoes a spin transition to a ferromagnet upon increasing $H$~\cite{eisenstein-PhysRevB.41.7910, clark-PhysRevLett.62.1536, shayegan-PhysRevB.45.3418, kukushkin-PhysRevLett.82.3665, Yacoby-natPhys2007}. 
% Most strikingly, the transition appears as a minimum of the activation gap in magnetoresistance experiments~\cite{eisenstein-PhysRevB.41.7910, clark-PhysRevLett.62.1536, shayegan-PhysRevB.45.3418}.
This behavior contrasts with {\em e.g.} the Laughlin fraction $1/3$ which is spin-polarized even in the limit $H=0$.

We focus on $\nu_{h,T} = \frac{2}{3}$. There, the phase transition in LL systems is understood as the transition from the bicomponent Halperin wavefunction (1 1 2)~\cite{Halperin83}, a spin singlet, to the particle-hole conjugate of the $1/3$ Laughlin state. This interpretation is supported by numerical evidence from exact diagonalization~\cite{xie-PhysRevB.40.3487, macdonald-PhysRevB.53.15845, wu-PhysRevLett.71.153, wu-PhysRevB.49.7515}.

In the flat-band limit, the main particularity of a Chern band compared to a Landau level is the non-uniform character of its Berry curvature and quantum metric. 
To study the effect of Berry curvature fluctuations on the fate of the spinful many-body ground state, we repeat our ED calculations for a range $0 \leq w_0/w_1 < 0.95$ of the relaxation parameter.
For $w_0/w_1 = 0.0$, we observe that the ground state is a spin singlet with an approximate threefold degeneracy and a gap to higher energy excitations. Upon adiabatic insertion of a magnetic flux $\varphi$ along one of the lattice axes, the three ground states flow into one another without closing the many-body gap (see Fig.~\ref{fig: singlet}a)); the original energy spectrum is restored after inserting three flux quanta. These numerical results strongly suggest the emergence of the Halperin (1 1 2)~\cite{Halperin83} state in a Chern band, in agreement with the expectation that Landau level physics should be restored in a Chern band with smooth enough Berry curvature. 
In this regime, the ferromagnetic FCI state may become the ground state upon addition of a Zeeman field.
Increasing the value of $w_0/w_1$, we observe a first order phase transition from the singlet to the FCI ferromagnet discussed in the previous paragraph (see Fig.~\ref{fig: singlet}b)) around $w_0/w_1 \simeq 0.7$. We observe a similar effect at $\nu_{h,T}=3/5$, although the small number (2) of sizes accessible to exact diagonalization prevents a proper extrapolation of the parameters of the phase transition (see Appendix B for further details).
For $w_0/w_1 \gtrapprox 0.9$, the ground state remains spin-polarized, but the many-body gap $\Delta E$ collapses. We interpret this phase transition as the result of the increase of correlation-induced dispersive terms due to quantum geometry fluctuations~\cite{lauchli-PhysRevLett.111.126802, grushin-PhysRevB.86.205125, repellin-2019arXiv190711723R}.

Our results, summarized in Fig.~\ref{fig: singlet}c), show a striking correlation between the root mean square $\tilde{F}$ of the Berry curvature distribution and the energy difference between the singlet and polarized ground state across all parameter values. This suggests that large $\tilde{F}$ favor spin-polarized states over spin unpolarized ones. We leave it to future work to examine the generality of this statement, and understand its theoretical origin.\\

\begin{figure}
\includegraphics[width=0.98\linewidth]{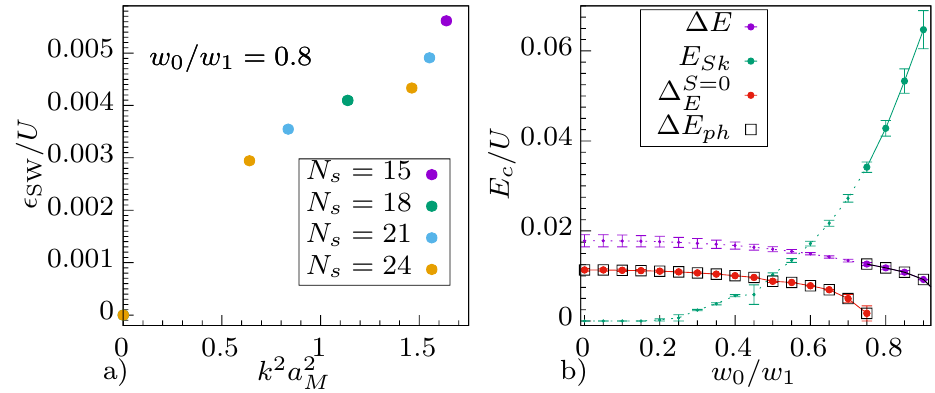}
\caption{a) Dispersion of the spin-wave excitation of the ferromagnetic $\nu_{h,T}=2/3$  FCI ground state at small $\mathbf k$. b) Energy of the three types of particle-hole excitations. For $w_0/w_1 < 0.75$, the ground state is a spin singlet with a charge gap $\Delta_E^{S=0}$. For $w_0/w_1 \geq 0.75$, the ground state is ferromagnetic. $\Delta E$ is the neutral gap of Fig.~\ref{fig: FM FCI}. The energy of a skyrmion particle-hole pair $E_{Sk}$ is extracted from a linear fit of $\epsilon_{\mathrm{SW}}(\mathbf{k})$ at small $\mathbf k$. Since $E_{Sk} > \Delta E$, skyrmions do not affect the activation gap $\Delta E_{ph}$. The values of $\Delta E$ and $E_{Sk}$ in the regime $w_0/w_1 < 0.75$ (dashed line) are obtained by restricting the calculation to the spin-polarized sectors. While in this regime they do not correspond to low energy excitations, we give them as an indication of the effect of $\tilde{F}$ on these quantities.}
\label{fig: charge gap}
\end{figure}

\textit{Activation gaps}
We now discuss the activation gap $\Delta E_{ph}$ -- lowest energy particle-hole excitation -- of the FCI states at $\nu_{h,T}=\frac{2}{3}$. $\Delta E_{ph}$ will determine the temperature where these phases may be observed in transport experiments. In the spin singlet FCI phase, we have found that the lowest energy excitation is a particle-hole excitation with spin $1$. 
In the ferromagnetic phase, two types of charged excitations may be considered. The first type is a particle-hole excitation within the spin-polarized sector; it is the many-body gap $\Delta E$ as shown in Fig.~\ref{fig: FM FCI}b) at $w_0/w_1=0.8$ and likely corresponds to the minimum of the magneto-roton mode~\cite{GMP1986, PhysRevB.90.045114}. Because of the $SU(2)$ spin symmetry in our system, we also need to consider the possibility of skyrmions. Skyrmions are long wavelength spin textures which can form as an excitation of a ferromagnetic FQH state~\cite{sondhi-PhysRevB.47.16419, kamilla-1996SSCom..99..289K, wojs-2002PhRvB..66d5323W, doretto-2005PhRvB..72c5341D, macdonald-PhysRevB.58.R10171}; in this case they have a fractional charge. In LL physics, skyrmions are the smallest charged excitation at small Zeeman energy~\cite{schmeller-PhysRevLett.75.4290, barrett-PhysRevLett.74.5112}. They could thus potentially be relevant to determine the activation of a ferromagnetic FCI in TBG/hBN.

Due to the large spatial extension of skyrmions, it is hard to evaluate their energy directly in finite-size calculations. However, it may be possible to calculate the spin wave dispersion of the ferromagnet, which at long wavelength is related~\cite{macdonald-PhysRevB.58.R10171} to the spin stiffness $\rho_s$
\begin{equation}
\epsilon_{\mathrm{SW}}(\mathbf{k}) = \frac{2\rho_s}{n}k^2
\end{equation}
where $n$ is the electronic density. The energy of a particle-hole skyrmion pair is then
\begin{equation}
E_{Sk} = 8\pi\rho_s
\end{equation}
Fig.~\ref{fig: charge gap}a) shows the spin-wave dispersion at small momentum $\mathbf{k}$, extracted from the exact diagonalization of Eq.~\ref{eq: screened coulomb} with $w_0/w_1 = 0.8$ in the partially polarized sector $S_z=N/2-1$. To obtain a sufficient number of points, we have used multiple finite cluster geometries for each system size (see Appendix~C for details). Finite-size effects prevent us from extracting a very precise value of $\rho_s$ from a linear fit, but we can still use this data to estimate the order of magnitude of $E_{Sk}$ and identify trends.
Fig.~\ref{fig: charge gap} shows the evolution of the energy of the three types of charged excitation with $w_0/w_1$. $E_{Sk}$ generically increases with the amplitude of Berry curvature fluctuations, similarly to what happens at integer filling~\cite{repellin-2019arXiv190711723R, chatterjee2019symmetry}. We find that in the regime where the ground state is ferromagnetic, the skyrmion excitations always have a higher energy $E_{Sk}$ than the spin-polarized particle-hole excitations, such that $\Delta E_{ph} = \Delta E$. For realistic values of the relaxation parameter $w_0/w_1 \simeq 0.8$, we find $\Delta E_{ph} \simeq 0.01U  \simeq 0.2 meV$.\\

\begin{figure}
\includegraphics[width = 0.98\linewidth]{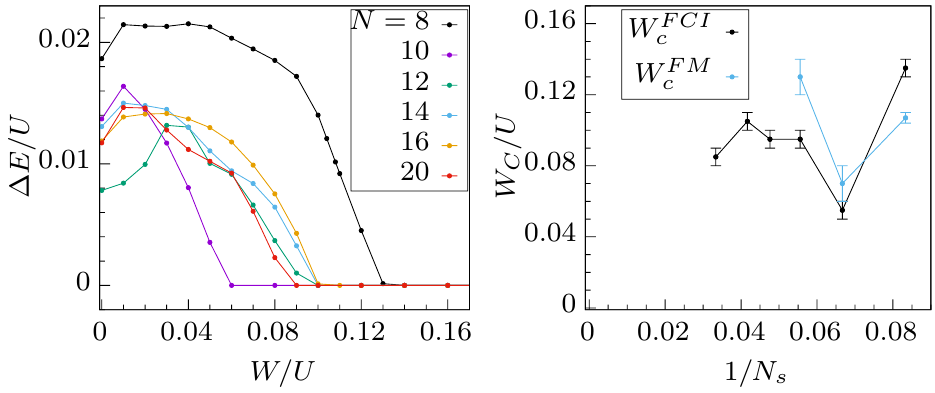}
\caption{a) Evolution of the many-body gap above the ferromagnetic FCI ground state upon adding a dispersion term of amplitude $W$ (Eq.~\eqref{eq: dispersion}) for $w_0/w_1 = 0.8$. b) Maximum value of the bandwidth to obtain a non-zero many-body gap ($W_c^{\mathrm{FCI}}$) or a fully polarized ground state ($W_c^{\mathrm{FM}}$).}
\label{fig: bandwidth}
\end{figure}

\textit{Effect of a finite bandwidth}
In the previous sections, we have shown the emergence of ferromagnetic FCI states in the limit where interactions are much larger than the bandwidth of the conduction band (flat band limit). Since this condition may not always be realized in experiments, we consider the addition of a kinetic term which gives the conduction band a width $W$.
\begin{equation}
H = H_V - \frac{W}{2}\sum_{\mathbf{k}}\cos\left( \mathbf{k}\cdot\mathbf{a_1} + \mathbf{k}\cdot\mathbf{a_2}\right) c^\dagger_{\mathbf{k}}c_{\mathbf{k}}
\label{eq: dispersion}
\end{equation}
where $\mathbf{a_1}, \mathbf{a_2}$ are the \moire lattice vectors, and we have taken a simplified dispersion (not the realistic one).

$W/U$ must be small enough to allow the ground state to be spin (and valley) polarized. This flat-band ferromagnetism -- familiar in the context of the FQHE -- was investigated quantitatively in the context of moir\'e narrow bands at integer filling~\cite{lee2019theory,bultinck2019anomalous, zhang2019bridging, zhang2019twisted, xie2018nature, alavirad-2019arXiv190713633A, repellin-2019arXiv190711723R}. Moreover, $W/U$ must be small enough to favor FCIs over metallic states in a Chern band. 
We find that these conditions are met roughly in the same regime $W < W_c^{\mathrm{FM}}\simeq W_c^{\mathrm{FCI}} \simeq 0.1U$. For $W > W_c^{\mathrm{FCI}}$, the many-body gap above the threefold ground state $\Delta E$ vanishes (see Fig.~\ref{fig: bandwidth}a)). For $W > W_c^{\mathrm{FM}}$ the ground state is no longer fully spin polarized (see Fig.~\ref{fig: bandwidth}b) for an extrapolation of the value of $W_c^{\mathrm{FM}}$ and $W_c^{\mathrm{FCI}}$ with system size). Note that $W_c^{\mathrm{FM}}$ is smaller, but of the same order of magnitude as its counterpart for an integer filling $\nu_T=1$ (for the same parameters, we have estimated $W_c^{\mathrm{FM}, \nu_T=3} \simeq 0.14$). Depending on the band structure parameters, the FCI ferromagnet may give way to either a ferromagnetic metal (due to large Berry fluctuations) or a spin unpolarized metal (due to a large bandwidth).\\

\textit{Discussion}
Our calculations demonstrate that fractional Chern insulators may be realized in TBG/hBN at fillings $\nu_T = \frac{10}{3}, \frac{17}{5}$. At these fillings valley polarization is seen in Ref.~\onlinecite{serlin2019intrinsic}. Motivated by this we studied models of TBG/hBN in the strong interaction limit, and assuming complete valley polarization. We did not assume spin polarization. For nominally realistic parameters of the single particle Hamiltonian, we nevertheless found that the ground state was spontaneously spin polarized and realizes a fractional Chern insulator state. An experimental signature of this state will be a quantized fractional anomalous Hall conductivity $\sigma_{xy} =  (4 - \nu_T) \frac{e^2}{h}$.  In the flat band limit we calculated an activation gap of $\approx 0.01 U \approx 0.2 meV \approx 2 K$ which, encouragingly, is within the reach of transport experiments.   The FQAH state remains stable for a range of non-zero bandwidth which we also estimated within a simple model. In practice such a non-zero bandwidth will reduce the activation gap. We may thus be cautiously optimistic that the FQAH state may be discovered in future studies of these fillings. 

A notable feature of our results is the close competition between spin singlet and spin polarized quantum Hall states even in the flat band limit, which we explored theoretically by varying the band structure parameter $\frac{w_0}{w_1}$. Though the spin polarized state is stabilized for the nominally realistic value $\frac{w_0}{w_1} \approx 0.8$, the spin singlet state is the ground state for a range of this parameter that includes the chiral limit $\frac{w_0}{w_1} = 0$. Given the uncertainties associated with the renormalized (as opposed to the bare) band parameters in the experimental system, it is important to keep open the possibility that any FQAH state observed at the fillings we have discussed may be a spin singlet. If such a spin singlet FQAH state is indeed present, it will have the same quantized Hall conductivity as the spin polarized state. However we expect that an in-plane magnetic field will drive (through the spin Zeeman coupling) a first order transition from the spin singlet to the spin polarized state which may reveal itself in experiments (much like in the analogous Landau level situation). 
Theoretically our work highlights a number of questions on the role of the quantum geometry of the band on flavor ordering in multicomponent fractional Chern insulators.  Moir\'e graphene systems provide a natural context where these issues arise, and we plan to return to these in the future.

\textit{Acknowledgements}
We thank Ya-Hui Zhang for insightful discussions and prior collaborations on related topics. We are also thankful to Andrea Young for useful discussions.
C.R. is supported by the Marie Sklodowska-Curie
program under EC Grant agreement No. 751859.
This work was supported by NSF grant DMR-1911666, and partially through a Simons Investigator Award from the Simons Foundation to Senthil Todadri. 
This work was supported in part by the NSF under Grant No. NSF PHY-1748958 through a stay of CR at the Kavli Institute for Theoretical Physics.

\textit{Note added}
During the final stages of preparing this manuscript, we became aware of two other preprints~\cite{2019arXiv191204907A, lewidth2019fractional} which discuss the existence of FCIs assuming full spin and valley polarization in TBG/hBN. Both the numerical results of Ref.~\onlinecite{2019arXiv191204907A} and the analytical approach of Ref.~\onlinecite{lewidth2019fractional} are in agreement with our findings where they overlap.

\bibliographystyle{apsrev4-1}
\bibliography{MoireFQH}

\hfill \break

\appendix

In this appendix, we provide additional numerical results to support the emergence of fractional quantum anomalous Hall effect in a model relevant for magic angle twisted bilayer graphene aligned with hBN (TBG/hBN).

\section{Particle Entanglement Spectrum of the $\nu_{h,T} = 2/3$ ground state}
\label{sec: PES}
For realistic band structure parameters (around lattice relaxation $w_0/w_1 = 0.8$), we have found a spin-polarized threefold almost degenerate ground state consistent with the particle-hole conjugate of the Laughlin $1/3$ state. We give here the details of the particle entanglement spectrum (PES)~\cite{sterdyniak-PhysRevLett.106.100405} analysis we have performed to confirm the topological nature of the ground state.

We start from the ground state $\Psi_{i = 0, 1, 2}$ of the system with $N$ electrons at filling fraction $\nu_{h,T} = 2/3$. $N_s$ is the number of points in the \moire Brillouin zone. 
As a model state, the Laughlin $1/3$ state has a PES with well known characteristics~\cite{sterdyniak-PhysRevLett.106.100405} which can be used to identify a FCI ground state~\cite{regnault-PhysRevX.1.021014}; to benefit from those, it is convenient to calculate the PES of the states $\tilde{\Psi}_i = \hat{\cal P} \Psi_i$ which are obtained through a particle-hole transformation (in the spin-polarized Hilbert space) of the three nearly degenerate ground states. $\tilde{\Psi}_i$ are wave functions of $\tilde{N} = N_s - N$ holes at filling fraction $1-\nu_{h, T} = 1/3$. 

\begin{figure}
\includegraphics[width=0.98\linewidth]{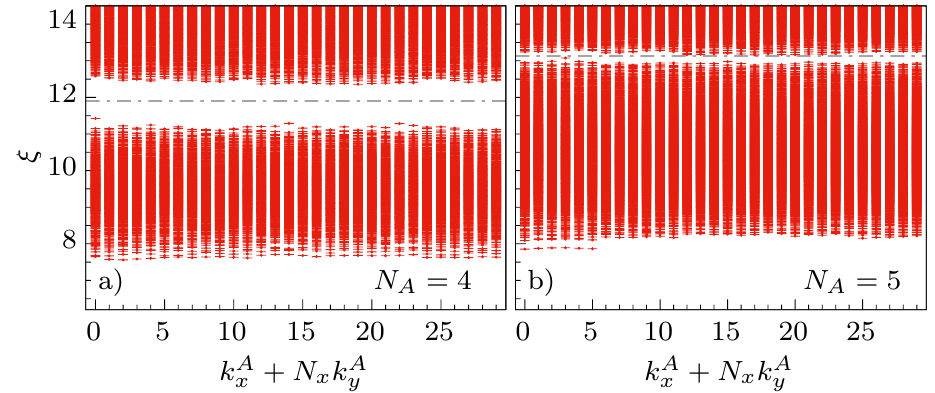}
\caption{PES of the spin-polarized ground state with $\tilde{N} = 10$ holes in a Brillouin zone with $N_x\times N_y=5\times 6$ points for $w_0/w_1=0.8$, for a subsystem with $N_A = 4$ and $N_A=5$. We use linearized coordinates for the total momentum $(K_{x,A},K_{y,A})$ in subsystem $A$. The black dashed line is guide to the eye indicating the position of the entanglement gap. The total number of levels below the gap is 9975 for $N_A=4$ (335 for even $K_{x,A}$ and 330 for odd $K_{x,A}$) and 23256 for $N_A=5$ (776 for $K_{y,A} = 0$, 775 otherwise). This matches the expectation for the Laughlin $1/3$ state.}
\label{fig: PES}
\end{figure}

We consider the density matrix $\rho = \frac{1}{3}\left(|\tilde{\Psi}_0\rangle\langle\tilde{\Psi}_0| + |\tilde{\Psi}_1\rangle\langle\tilde{\Psi}_1| + |\tilde{\Psi}_2\rangle\langle\tilde{\Psi}_2|\right)$. We numerically calculate the reduced density matrix $\rho_A = \Tr_B \rho$, obtained by tracing over $N_B \equiv \tilde{N} - N_A$ fermions. Because the partition of the system leaves its geometry intact, $\rho_A$ commutes with the translation operators along the two axes of the lattice. Thus, we can label the eigenvalues $\xi$ of $-\ln \rho_A$ with the total momentum $(K_{x,A},K_{y,A})$ in subsystem $A$. In Fig.~\ref{fig: PES}, we give the PES of the spin-polarized ground state with $\tilde{N} = 10$ holes in a Brillouin zone with $5\times 6$ points for $w_0/w_1=0.8$ for $N_A = 4$ and $N_A=5$. We observe a clear entanglement gap. The number of states below the gap in each $(K_{x,A},K_{y,A})$ sector matches the number of quasihole states in the FQH system with $N_A$ fermions and $N_s$ flux quanta, as predicted using the generalized exclusion principle~\cite{haldane-PhysRevLett.67.937} and the FQH to FCI momentum mapping~\cite{bernevig-PhysRevB.85.075128}.

\section{Numerical evidence for a FCI at $\nu{h,T} = 3/5$ and $w_0/w_1 = 0.8$}

As mentioned in the main text, for $\nu_{h,T} = 3/5$ and realistic values of the relaxation parameter $w_0/w_1$, our results are consistent with a spin-polarized FCI ground state. The spin polarization is observed in systems with $N_s=10$ and $N_s=15$ (respectively $N = 6$ and $N=9$), the largest systems which we can simulate without assuming any spin polarization.

\begin{figure}
\includegraphics[width=0.98\linewidth]{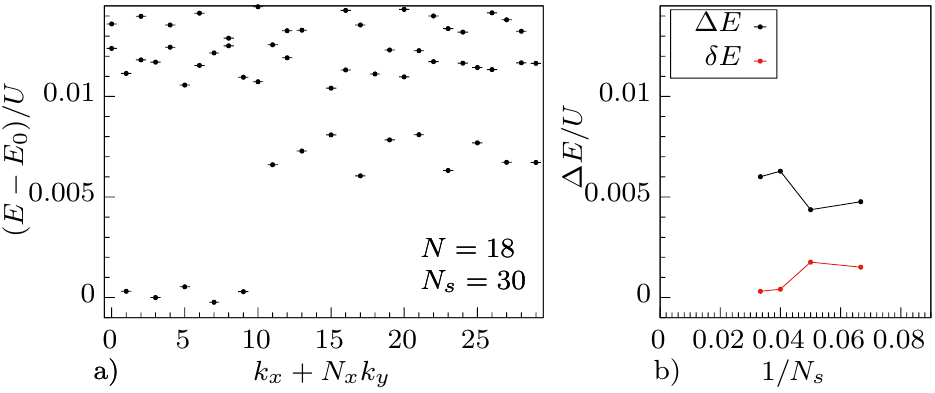}
\caption{Evidence for a fractional Chern insulator ferromagnet at $\nu_{h, T} = 3/5$. a)~Low-energy spectrum of the system with $N=18$ fermions and $N_s = 30$ moir\'e sites, for $w_0/w_1 = 0.8$, in the sector $S_z=N/2$. b)~Finite-size extrapolation of the many-body gap $\Delta E$ and ground state degeneracy splitting $\delta E$ for $w_0/w_1 = 0.8$.}
\label{fig: 3/5 spinless}
\end{figure}

Assuming spin-polarization in larger systems, we observe a gapped ground state with an approximate $5$-fold degeneracy. This is consistent with a FCI with the same topological order as the particle-hole conjugate of the $2/5$ composite fermion FQH state~\cite{jain-PhysRevLett.63.199, liu-PhysRevB.87.205136, lauchli-PhysRevLett.111.126802} (Fig.~\ref{fig: 3/5 spinless}a shows the low-energy spectrum for $N=18$). We also verified that the momentum quantum number of the ground state manifold can be predicted from the momentum of the degenerate FQH ground state at $\nu=3/5$, using the mapping of Ref.~\onlinecite{bernevig-PhysRevB.85.075128}. Fig.~\ref{fig: 3/5 spinless}b) shows the extrapolation of the many-body gap $\Delta E$ with system size, as well as the degeracy splitting $\delta E$.

\begin{figure}
\includegraphics[width=0.98\linewidth]{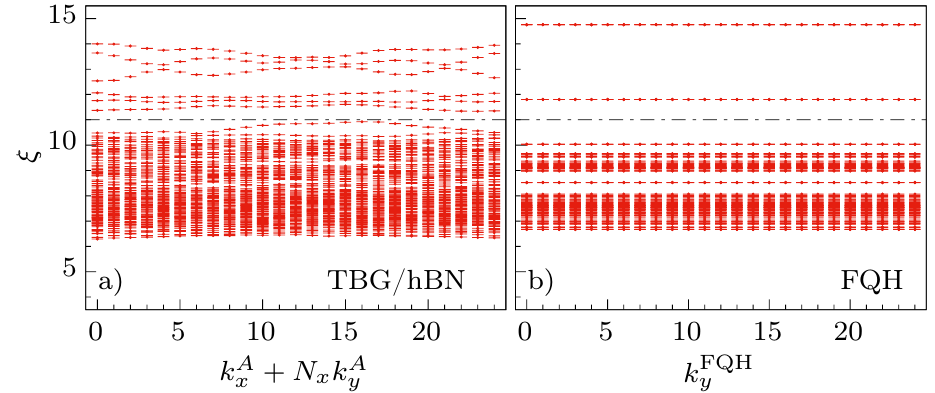}
\caption{a) PES of the $\nu_{h,T} = 3/5$ spin-polarized ground state manifold with $N=15$ and $N_s=25$, after particle-hole conjugation in the spin-polarized Hilbert space, for $w_0/w_1=0.8$, and a subsystem with $N_A = 3$. We use linearized coordinates for the total momentum $(K_{x,A},K_{y,A})$ in subsystem $A$. b) PES of the $\nu=2/5$ FQH fivefold degenerate ground state for $N=10$, $N_{\phi} = 25$ flux quanta and $N_A = 3$. The number of states below the black dashed line is the same in both systems.
}
\label{fig: PES 3/5}
\end{figure}

We calculated the PES of the ground state manifold (after particle-hole conjugation to obtain a state at filling fraction $2/5$, similarly to Section~\ref{sec: PES}). We compared the PES in TBG/hBN to the PES of the five degenerate ground states obtained by exact diagonalization of the hollow-core two-body interaction ($V_1$ pseudopotential) at $\nu=2/5$. These two PES are shown for $N_s=25$ and $N_A=3$ in Fig.~\ref{fig: PES 3/5}; we find an entanglement gap with the same number of states below the gap for both TBG/hBN and FQH models. The same method was used in Ref.~\onlinecite{liu-PhysRevB.87.205136} to demonstrate the existence of $2/5$ FCIs in a toy-model.

Finally, we probed the role of band geometry by changing the value of the relaxation parameter $w_0/w_1$. We find a phase transition from a spin unpolarized (at small values of $w_0/w_1$) to a spin polarized ground state. With only two available system sizes, it is hard to extrapolate the position of the transition; still, we find that it occurs around $w_0/w_1 \simeq 0.75$ for both $N_s=10$ and $N_s=15$. As expected for large Berry curvature fluctuations, we find that the neutral gap $\Delta E$ above the fivefold quasi-degenerate ground state vanishes around $w_0/w_1 \simeq 1$.\\

\section{Geometry of the finite-size clusters}
For our exact diagonalization calculations, we have used finite-size clusters with periodic boundary conditions. The finite size leads to a discretization of the Brillouin zone into $N_s$ allowed values of the total momentum $\mathbf{k}$. 
The \moire lattice is spanned by two vectors $\mathbf{a_1},\mathbf{a_2}$. The periodic boundary conditions are defined by two vectors $\mathbf{d_1}, \mathbf{d_2}$, which are linear combinations with integer coefficients of $\mathbf{a_1},\mathbf{a_2}$. While $\mathbf{a_1},\mathbf{a_2}$ are fixed, there are several possible choices for $\mathbf{d_1}, \mathbf{d_2}$ for a given system size $N_s$ (see Appendix~A of Ref.~\onlinecite{repellin-PhysRevB.90.245401} for more details). We define the aspect ratio as $\kappa = |\mathbf{d_1}|/|\mathbf{d_2}|$, and call $\alpha$ the angle between $\mathbf{d_1}$ and $\mathbf{d_2}$.

For the small systems accessible to exact diagonalization, the geometry of the cluster may play an important role. Indeed, the emergence of FCIs was shown to be facilitated~\cite{regnault-PhysRevX.1.021014, lauchli-PhysRevLett.111.126802, bernevig-2012arXiv1204.5682B} if $\kappa \simeq 1$.
For larger systems, we expect that the geometry does not influence the stability of the FCI as long as the length $l_1, l_2$ of each cycle of the torus is large compared to the correlation length of the ground state ($l_i = |\mathbf{d_i}|$ when $\alpha  =\pi/2$).
In finite-size, it thus makes sense to choose a cluster geometry which maximizes $\mathrm{min}(l_1, l_2)$.
To do so, we have chosen geometries with an aspect ratio $2/3 \leq \kappa \leq 3/2$ and an angle $\pi / 3 \leq \alpha \leq 2\pi/3$ for all of our exact diagonalization calculations.
The geometry choice is especially important to calculate the spin-wave dispersion of the FCI ferromagnet (Fig.~4a of the main text).
Indeed, the discretization of $\mathbf{k}$ is especially severe for the small systems we can reach with exact diagonalization; it sets a lower bound on the allowed values of $|\mathbf{k}| \rightarrow 0$. We have taken advantage of the freedom in the choice of $\mathbf{d_1}, \mathbf{d_2}$ to obtain several values of $|\mathbf{k}|$ in the range $|\mathbf{k}|^2 a_M^{2}< 1.75$ where $a_M$ is the \moire lattice constant.

\hfill \break

\end{document}